\documentclass[nofootinbib,aps,tightenlines,amssymb,letterpaper]{revtex4}

\newcommand{\f}{\begin{equation}}
\newcommand{\ff}{\end{equation}}
\newcommand{\be}{\begin{equation}}
\newcommand{\ee}{\end{equation}}
\newcommand{\beq}{\begin{eqnarray}}
\newcommand{\eeq}{\end{eqnarray}}
\newcommand{\dr}{\partial}

\newcommand{\e}{\epsilon}

\begin{document}

\begin{titlepage}
\title{\LARGE The linearization of the Kodama state}
\author{ Laurent Freidel}
\email{lfreidel@perimeterinstitute.ca}
\affiliation{\vspace{1mm}Perimeter Institute for Theoretical
Physics\\ 35 King street North, Waterloo  N2J-2G9,Ontario,
Canada\\} \affiliation{Laboratoire de Physique, \'Ecole Normale
Sup{\'e}rieure de Lyon \\ 46 all{\'e}e d'Italie, 69364 Lyon Cedex
07, France }\thanks{UMR 5672
du CNRS}

\author{Lee Smolin}
\email{lsmolin@perimeterinstitute.ca}
\affiliation{\vspace{1mm}Perimeter Institute for Theoretical
Physics\\ 35 King street North, Waterloo  N2J-2G9,Ontario,
Canada\\}
\date{\today}

\begin{abstract}
    We study the question of whether the linearization of the Kodama state
    around classical deSitter spacetime is
    normalizable in the inner product of the theory of linearized
    gravitons on deSitter spacetime.  We find the answer is no in the
    Lorentzian theory. However, in the Euclidean theory the corresponding
    linearized Kodama state is delta-functional normalizable.
    We discuss whether this result invalidates the conjecture that the
    full Kodama state is a good physical state for quantum gravity with
    positive cosmological constant.
\end{abstract}
\maketitle
\end{titlepage}

\section{Introduction}

A central problem of loop quantum gravity is to show whether or
not the theory has a good low energy limit \cite{reviews}. This
must reproduce as an appropriate approximation, classical general
relativity and quantum field theory on fixed backgrounds. This is
a hard problem for the same reason it is hard to derive the
properties of liquids or solids from first principles given only
the quantum theory of atoms.  Even if a particular choice of
hamiltonian constraint or spin foam amplitude enjoys all the
properties that could be asked from a fundamental point of view,
it is far from trivial to derive the low energy behavior.  An
infinite number of exact solutions to all the constraints are
known, but little is known about the low energy description of
most of them.  A number of directions are under development to do
address this question, based on extensions to quantum gravity of
coherent states \cite{coherent} or the renormalization group
\cite{rg}.

A possibly important piece of evidence for this question is the
Kodama state \cite{ Kodama:1990sc,Kodama:1988yf}. This is the, so
far, unique, precisely known, quantum state of the gravitational
field that has both an exact description at the Planck scale {\it
and} a semiclassical description. Furthermore, it exists only for
nonzero cosmological constant, $\Lambda$, which supports the
intuition that the cosmological constant is an essential parameter
of any quantum theory of gravity.

The understanding of the Kodama state as a semiclassical description is
straightforward\footnote{More details are contained in a
pedagogical introduction to the Kodama state see \cite{Smolin:2002sz}.}.
In the Ashtekar formulation, the configuration variable of
the gravitational field is the self-dual connection, $A_{a}^{i}$.
This is a complex variable for the Lorentzian theory, so it is valued
in the complexification of $SU(2)$. This fact will be important for
what follows.

To construct a semiclassical description of deSitter
spacetime\footnote{We work in this paper with the case of
$\Lambda>0$, but most results extend also to negative $\Lambda$.}
we require a Hamilton-Jacobi functional $S(A)$ with the property
that deSitter spacetime is one of its trajectories. A convenient
way to describe deSitter spacetime in the language of the Ashtekar
formalism is that it is the unique Lorentzian self-dual spacetime.
The self-dual condition is expressed as \be B^{a}_{i} = {1\over 2}
\epsilon^{abc}F_{bc}^{i} = {\Lambda \over 3} E^{a}_{i}, \ee
$E^{a}_{i}$ is the densitized spatial triad (or, equivalently the
pull back to the spatial slice of the self-dual two form) and the
left-handed components of the spacetime curvature are given by \be
F_{ab}{}^{i} = \dr_{a}A_{b}{}^{i}- \dr_{b}A_{a}{}^{i} +
\epsilon^{i}_{jk} A_{a}{}^i A_{b}{}^k \ee

$E^{a}_{i}$ is canonically conjugate to $A_{a}^{i}$ (up to a factor of
$\imath G$). Thus, it must be true that,
\be
E^{ai}= \imath G {\partial S(A) \over \partial A_{ai}} = {3 \over
\Lambda} \epsilon^{abc}F_{bc}^{i}
\ee
There is, up to a constant, a single unique solution to
this equation, which is
\f
S(A)= -\imath{ 3\over G \Lambda } \int_{\Sigma} Y(A)_{CS}
\ff
where $Y(A)_{CS}$ is the Chern-Simons three form
\be
Y(A)_{CS}=\frac{1}{2}A \wedge dA + \frac{1}{3}A\wedge A\wedge A.
\ee
Thus, the semiclassical state for deSitter must be the Kodama
state
\f
\Psi_{K}(A) = e^{{3 \over  \lambda } \int_{\Sigma} Y(A)_{CS}}
\ff
where $\lambda=G\hbar \Lambda$ is the dimensionless cosmological
constant.  We note there is no $\imath$ in the exponent because
\f
\hat{E}^{ai}= \hbar G {\partial \over \partial A_{ai}}
\label{Edef}
\ff

It is then straightforward to use the Born-Oppenheimer method to
develop a semiclassical expansion around the Kodama state.
This is described in \cite{Smolin:2002sz,Smolin:1994qb,Soo:1993an} 
and applications to cosmology
are described in \cite{Alexander:2003wb}.

However, the Kodama state has other properties that suggest it
may be more than just a semiclassical approximation to the quantum
state of the gravitational field. First, it is in fact an
exact solution to the constraints of quantum gravity. This is easy to
see naively, but it is also true when consideration is given to
issues of regularization and operator ordering.

Also, in the case of the Euclidean theory all the properties
described so far hold, with all the $\imath$'s removed. In that
case there are many self-dual solutions, so the Kodama state can
be the basis for a semiclassical expansion around any of them. The
Euclidean theory is simpler in several respects, one of them is
that the connection $A_{a}^{i}$ is real, so it is valued in the
real form of $SU(2)$. Because of this, the Euclideanized Kodama
state is a phase, 
\f 
\Psi_{Eucl-K}(A) = e^{{3 \imath \over
\lambda } \int_{\Sigma} Y(A)_{CS}} 
\label{purephase}
\ff 
because there is now an $\imath$ in (\ref{Edef}). 
Moreover, the loop transform of the
Euclidean Kodama state, defined by 
\f 
\tilde{\Psi}_{Eucl-K}(\Gamma ) = \int d\mu(A) T[\Gamma, A] 
\Psi_{K}(A) , 
\ff 
where $T[\Gamma, A]$
is the traced holonomy of the spin network $\Gamma$ and $d \mu(A)$ is an
appropriate measure,  is well understood and has a striking property: It
is proportional to the Kauffman bracket for framed quantum spin
networks with a unimodular quantum group parameter. In the case of
the Lorentzian Kodama state the issue is more subtle, in order to
define the Loop transform we have to take the integration contour
along a real section of $SL(2,C)$ connections defined by the
reality conditions. There is no rigorous proof on what the result
should be. However if we assume that we can deform the contour of
integration to be over $SU(2)$ connection, and disregards
convergence issue we expect the result to also be proportional to
the Kauffman bracket with real quantum deformation parameter.

The purpose of this paper is to address several questions that
have been raised concerning the Kodama state \cite{Witten:2003mb}.
One of them is that a similar state exists in the case of $SU(2)$
Yang-Mills theory\cite{jackiw}.
However, in that case it is believed to be unphysical, for the
following reasons, which we now discuss.

In the $SU(2)$ Yang-Mills theory (in Lorentzian signature)  
the analogue to the Kodama state is the solution to
$[E^{a}_{i} - \imath B^{a}_{i}]\Psi (A)=0$, where 
$E^{a}_{i} = \imath g^{2} \delta / \delta A_a^{i}$,
where $g$ is the Yang-mills coupling.
The solution is real, and is given by
\f
\Psi (A) = e^{{1 \over \hbar g^{2} } \int_{\Sigma}  Y(A)_{CS}}
\ff
This state is not 
normalizable.  Because the theory is not diffeomorphism
invariant, and there is no Hamiltonian constraint, the physical norm is
\f
<\Psi |\Psi>_{YM} = \int dA |\Psi (A)|^{2}
\label{YMIP}
\ff
This tells us that in Yang-Mills theory
\f
<K |K>_{YM} = \int dA
e^{{2 \over \hbar g^{2} } \int_{\Sigma}  Y(A)_{CS}},
\ff
This is not finite
because there are directions in the configuration space which are
unbounded.

Does this argument extend to quantum gravity?  It is easy to see
that it does not directly. First of all, the physical inner
product is not given by (\ref{YMIP}).  This is what is called the
``kinematical inner product'' as it defines the kinematical
Hilbert space, which is the arena for the definition of the
quantum constraints. But physical states live in the subspace of
states annihilated by the diffeomorphism and hamiltonian
constraints. This subspace should be given an inner product
structure defined by the physical reality conditions. This is that
real physical observables (meaning observables that commute with
all the constraints) must be represented by Hermitian operators.
The physical states are then not expected to  be normalizable in
the kinematical inner product.

It is unfortunate that there is not known a closed form expression
for the physical inner product in loop quantum gravity, in either
the Euclidean or the Lorentzian theory. The physical inner product
can be, however, expressed as a path integral and, in loop quantum
gravity, there are explicit proposals for the physical inner
product in terms of the spin foam formalism \cite{spinfoam}.
Recent convergence results on spin foam amplitudes show that the
projection operator will be ultraviolet finite \cite{convergence}
and the fact that the representation theory for non-zero
cosmological constant is q-deformed implies that the spin foam
summations are also infrared finite \cite{qdef-foams}.
Unfortunately, it is so far not possible to use this to test
whether or not the Kodama state is normalizable with respect to
the physical inner product.

We can ask a simpler question: is the Kodama state  
normalizable in the kinematical inner product. In the Euclidean 
case it is delta-functional normalizable, because the state
(\ref{purephase}) 
is a pure phase. 
In the Lorentzian case the situation is more complicated than
in Yang-Mills theory because  $A_{a}^{i}$ is
a complex variable, so the inner product must be defined by a choice
of a contour. Furthermore, $S(A)$ is a complex function, so the
required integral has the form
\f
<K|K> = \int d\mu (A) e^{{3 \over  \lambda } \int_{\Sigma}
2{\cal R}e  Y(A)_{CS} } \Delta(A,\bar{A})
\ff
The contour is discussed in \cite{Soo:2001qf,Paternoga:2000cb},
and $\Delta$ is a functional implementing the projection on the kernel of
the hamiltonian constraint,
but the result is that it is not
presently known whether this expression is convergent or divergent.

There is another question we can ask, which will be the subject of
the rest of this paper.  We can linearize classical general
relativity on deSitter spacetime. This gives us a classical theory
of tensor fields, $a_{ab}$ propagating linearly on deSitter
spacetime. We can quantize this theory. This can of course be done
both in the usual ADM variables and in the Ashtekar variables.
The result is the quantum theory of linearized gravitons on
deSitter spacetime.  This is not hard to do, following previous
results \cite{Ashtekar:mz,Ashtekar:vz} in the linearized Ashtekar
formalism at $\Lambda=0$, and we will carry it out in the next
section.

In the course of constructing this
theory one solves the linearization versions of the constraints to
find the physical Hilbert space of the linearized theory, which we
may call ${\cal H}_{linear}$.  It has an inner product which
can be found exactly by solving the linearization of the reality
conditions.  One can also linearize the Kodama state, arriving, for
the Lorentzian case, at
\f
\Psi_{LK}(a) = e^{{3 \over  \lambda } \int_{\Sigma} Y(a)_{LCS}}
\ff
where $Y(a)_{LCS}$ is the quadratic truncation of the Chern-Simons
three form.  This state is annihilated by the linearized quantum
constraints. So it is a functional of the same variables that
linearized quantum states in ${\cal H}_{linear}$ depend on. (These are not
surprisingly the symmetric, trace-free, transverse components of
the connection.)

One can then ask if $\Psi_{LK}(a)$ is normalizable or
even delta-functional normalisable in the inner product
of the linearized theory.  The answer, as we show in the next two
sections is no, for the Lorentzian case.  In the Euclidean case however,
the linearized Euclidean Kodama state is
delta-functional normalizable in the inner product of the linearized
Euclidean theory.
In the next  sections we carry out the construction of the
linearized theory and compute the norm of the linearized Kodama state.
In the last section we discuss the implications of the results found.

\section{notations and definitions}

We denote by $A$ the restriction of the self dual connection to a
3 dimensional spacelike slice.
The phase space
variables are pairs
  $(A_a^i, E^b_j)$
  where A is a self dual connection and E is the
densitized inverse frame field. This means that
$ E^a_i/\sqrt{\det(E)}=e^a_{i}$ is the inverse frame field.
The self dual
connection can be expressed in terms of the usual geometrical
variables as follows
\be
A_{a}^i = \Gamma_{a}^i + i K_{ab}e^b_{i}
\ee
where $K_{ab}$ is the extrinsic curvature tensor, and $\Gamma$
is the spin connection satisfying
\be
de^i + \epsilon^i _{jk} \Gamma^j \wedge e^k=0,
\ee
where $e^i=e^i_a dx^a $ is the frame field.
From this relation it is clear that the
quantum commutation relations are
\be
 [A_a^i(x), E^b_j(y)]=l_P^2 \delta_a^b \delta^i_j\delta(x-y),
 \ee
where $l_P = \sqrt{\hbar G}$ is the Planck length and the reality conditions are given by
 \beq
 \label{real}
 \bar{E}^a_i&=&E^a_i \\
 A_{a}^i +\bar{A}_{a}^i &=& 2 \Gamma_{a}^i(E).
 \eeq
 There are three types of constraints.
 The gauge constraint
 \be
 G^i= \nabla_{a}E^a_{i}= \dr_{a}E^a_{i} + \epsilon^{ijk}
 A_{aj} E^a_{k}=0,
 \ee
 the diffeomorphism constraint
 \be
 D_{a} = E^b _{i}F_{ab}^i=0,
 \ee
 and the hamiltonian constraint
 \be\label{H}
 C(x)= \frac{1}{2\sqrt{\det(E)}}
 \e_{abc}\e^{ijk}E^a_{i} E^b_{j} (B^c_{k}+ \mathrm{H}^{2} E^c_{k})=0,
 \ee
 where the magnetic field is defined by
 $\e_{abc} B^c_{k} =F_{ab}^k$ and we have introduced a constant $\mathrm{H}$
 (the Hubble constant)
 which is related to the cosmological constant $\Lambda$ by
 \be
 \mathrm{H}^{2} = \frac{\Lambda}{3}.
 \ee
 Note that all the constraints are weight one density.
 One particular set of solutions are the `self dual'
 solutions for which the electric field is proportional to the
 magnetic field
 \be\label{selfdual}
 B^c_{k}+ \frac{\Lambda}{3} E^c_{k}=0.
 \ee
 de Sitter space is a self dual solution, this can be seen by
 choosing a particular gauge where $E$ is flat
 \be
 A_{a}^i = i  f \mathrm{H} \delta_{a}^i, \, E^a_{i}= f^2 \delta^a_{i}.
 \ee
 $f$ labels gauge equivalent solutions, it is
 related to the usual
 time of inflationary coordinates.
 \be\label{flatslice}
 f =\exp(+\mathrm{H} t),
 \ee
\be
ds^2 = -dt^2 + e^{2\mathrm{H} t} d\vec{x}^2.
\ee
Note that $\eta= (Hf)^{-1}$ is the conformal time in flat slicing
\be\label{confslice}
ds^2=\frac{1}{\mathrm{H}^2\eta^2}(-d\eta^2 +  d\vec{x}^2).
\ee

\subsection{linearized gravity}

We consider a perturbation of dS background
\beq
A_{a}^i &=& i \mathrm{H} f \delta_{a}^i + a_{a}^i \label{aexp}\\
E^a_{i} &=&   f^2 \delta^a_{i} +e^a_{i}\label{line}
\eeq
The constraints of gravity linearized around de Sitter background  are
 the gauge constraint
\be
 g^i= \dr_{a}e^{ai} + i \mathrm{H} \epsilon^{iak}
  e_{ak} +  \epsilon^{ija} a_{aj} =0,
\ee
 the diffeomorphism constraint
\be
d_{a}=f^{2}(\dr_{b}a_{a}^b -\dr_{a}a_{b}^b  +i\mathrm{H} f \epsilon_{a}{}^{bc}
a_{bc} +\mathrm{H}^{2}\epsilon_{a}{}^{bc} e_{bc})=0,
\ee
 and the hamiltonian constraint
\be
c={f} \epsilon^{abc}\dr_{a} a_{bc} + 2i \mathrm{H} f^{2} a_b^b
+\mathrm{H}^{2}f e_{a}^a =0.
\ee
The gauge fixing conditions we choose are such that
$\dr_{a}a_{b}^a=0$ to gauge fix the gauge constraint and  $e^{[ab]}=0 $
in order to gauge fix the diffeomorphism constraint.
Together with $a_{b}^b =0$ for the hamiltonian constraint.
This means that the zero mode part of $a_{b}^b $ is a good time variable.
This is clear since  $A_{b}^b = 3i\mathrm{H} f + a_{b}^b $ so a non zero
constant $a_{b}^b $ can be understood as a shift in time
\be
\delta f = \frac{a_{b}^b }{3i\mathrm{H}}.
\ee
Moreover it is easy to see that
 $\partial_t f=\{\delta f, \int N c \}= N(x)\mathrm{H} f$.
 So if we choose $N=1$ (resp. $N=f$) we recover the parametrization
 (\ref{flatslice})( resp. \ref{confslice})
  of $f$ in terms of inflationary time (resp. conformal time).

Overall, this means that both $a$ and $e$ can be taken to be symmetric,
transverse and
traceless tensors.
Such fields carry two degree of freedom per spacetime points, the positive
and negative helicity.
In order to describe these degree of freedom it is convenient to work in
momentum space.
In this space we introduce for each momentum $k$ a basis $ m^a(k),
\overline{m}^a(k), k^a$
($m,\overline{m}$ being complex conjugate) satisfying
\be
k^a m_a(k) =0, m_a(k) m^a(k)=0, \overline{m}_a(k) m^a(k) =1
\ee
and
\be
\epsilon_{abc} ik^a m^b(k) =|k| m^c(k).
\ee
these definitions imply that $m^a(-k) =\overline{m}^a(k)$.
The symmetric traceless transverse fields $a_{ab}(x), e_{ab}(x)$ can be 
expressed in terms of positive
and negative helicity fields
$a^\pm(k),e^\pm(k)$
\be
a_{ab}(x)= \int \frac{d^3k}{(2\pi)^{3/2}} e^{i\vec{k}\cdot\vec{x}}(
a^+(k)m_a(k)m_b(k) + a^-(k)\overline{m}_a(k) \overline{m}_b(k),
\ee
\be
e_{ab}(x)= \int \frac{d^3k}{(2\pi)^{3/2}} e^{i\vec{k}\cdot\vec{x}}(
e^+(k)m_a(k)m_b(k) + e^-(k)\overline{m}_a(k) \overline{m}_b(k).
\ee

\subsection{linearized reality conditions and commutation relations}

The commutation relations in terms of the helicity fields are given by
\be
[a^{\pm}(k),e^{\pm}(p)]= l_P^2\delta^3(k+p); [a^\pm(k),e^\mp(p)]=0.
\ee
The linearized spin connection associated with the linearized $E$
field (\ref{line}) is given by
\be
\gamma_a^i = -f^{-2} \epsilon^{ijk}(\partial_je_{(ka)} +\cdots ),
\ee
so when $e$ is symmetric and traceless
\be
\gamma_a^i = -f^{-2} \epsilon^{ijk}\partial_je_{(ka)}.
\ee
The linearized reality conditions are therefore given by
\beq \label{linreal}
\bar{e}^\pm(k) & = &e^\pm(-k), \\
a^+(k) +\bar{a}^+(-k) &=& - 2 f^{-2}|k| e^+(k),\\
a^-(k) +\bar{a}^-(-k) &=& + 2 f^{-2}|k| e^-(k). \eeq The reality
condition on $e$ is a consequence of the other ones when $k\neq
0$. Also, it is important to realize that there is a sign
difference between the two helicities. The commutators between the
self dual connection and its complex conjugate are given by \beq
\label{ancre}
[ \bar{a}^+(k), a^+(p) ] & = & +2l_P^2 f^{-2}{|k|} \delta^3(k-p), \\
{[} \bar{a}^-(k), a^-(p){]}  & = & -{2l_P^2 {f^{-2}}|k|} \delta^3(k-p), \\
{[} \bar{a}^-(k), a^+(p) {]} & = & 0. \eeq This clearly shows that
$a^+$ is a creation operator (raising energy) whereas $a^{-}$ is
an annihilation operator (lowering energy).

\section{physical states}
At the quantum level we decide to work in the polarization where the
self dual connection $A$ is diagonal. The  physical states are solutions
of the linear
constraints and we have seen that we can choose the gauge fixing such
that the wave function depends only on the symmetric transverse traceless
part of the perturbation  so in this polarization the wave function
$\psi$ is a functional of $(z^+(k),z^-(k))$ and
  \be
  a^\pm(k) \psi(z^+,z^-)= z^\pm(k) \psi(z^+,z^-).
  \ee
The representation we work with is the one for which
the frame fields acts as a derivative operator
  \be
  e^\pm(-k) \psi(z^+,z^-) =
  -l_{P}^{2} \frac{\partial \psi(z^+,z^-)}{\partial z^\pm(k)}.
  \ee
Given this representation of the commutator algebra the scalar product
is now uniquely determined by the reality conditions.
The scalar product is expressible as an infinite dimensional integral
\be
<\phi|\psi> = \int {\cal{D}}^2z^+ {\cal{D}}^2z^- \overline{\phi}
(\bar{z}^+,\bar{z}^-) e^{F(z^\pm,\bar{z}^\pm)} \psi(z^+,z^-)
\ee
where $D^2z = \prod_k dz(k) d\bar{z}(k)$ is the usual path integral measure
and the functional $F$ satisfies
the following reality conditions
\beq
\frac{\partial F}{\partial \bar{z}^+(k)}=
 \frac{\partial F}{\partial {z}^+(-k)}=
 -\frac{f^2}{2 |k|l_{P}^{2} }(z^+(k) +\bar{z}^+(-k)),\\
 \frac{\partial F}{\partial \bar{z}^-(k)}=
 \frac{\partial F}{\partial {z}^-(-k)}=
 +\frac{f^2}{2 |k|l_{P}^{2} }(z^-(k) +\bar{z}^-(-k)).
 \eeq
So $F$ can be written as $F =F^+(z^+,\bar{z}^+) + F^-(z^-,\bar{z}^-)$ where
\be
F^\pm(z^\pm,\bar{z}^\pm)=\mp f^2 \int \frac{d^3k}{4|k|l_{P}^{2}}(z^\pm(k) +
\bar{z}^\pm(-k))(z^\pm(-k) +\bar{z}^\pm(+k)).
\ee

The representation we just described is equivalent for the positive helicity
to the usual Bargmann-Fock coherent state representation.
If we define
\be \phi_B(z^+) \equiv e^{- f^2 \int \frac{d^3k}{4|k|l_{P}^{2}}z^+(k)z^+(-k)}
\phi(z^+)\ee
The scalar product in term of these functionals is
\be
<\psi_B|\phi_B> = \int {\cal{D}}^2z^+ \bar{\psi}_B(\bar{z}^+) e^{- f^2 \int
\frac{d^3k}{2|k|l_{P}^{2}}z^+(k)\bar{z}^+(k)} \phi_B(z^+).
\ee
The Fock vaccua is annihilated by $\bar{a}^+$ and  given by the functional
$\psi^{(0)}_B(z^+)=1$.
This state is normalizable in the sense that each mode $k$ is normalizable,
the normalization being
\be
\int d^2z(k) e^{-\frac{f^2 |z(k)|^2}{2|k|l_{P}^{2}}}= 
\frac{2\pi|k|l_{P}^{2}}{f^2}.
\ee

For the negative helicity we can get an analogous description by
defining \be \phi_{\bar{B}}(z^+) \equiv e^{+ f^2 \int
\frac{d^3k}{4|k|l_{P}^{2}}z^-(k)z^-(-k)}\phi(z^-)\ee The scalar
product being \be <\psi_{\bar{B}}|\phi_{\bar{B}}> = \int
{\cal{D}}^2z^+ \bar{\psi}_{\bar{B}}(\bar{z}^+) e^{+f^2 \int
\frac{d^3k}{2|k|l_{P}^{2}}z^+(k)\bar{z}^+(k)} \phi_{\bar{B}}(z^+).
\ee The state annihilated by $\bar{a}^-$ is also  given by the
functional $\psi^{(0)}_{\bar{B}}(z^-)=1$. However due to the wrong
sign in the exponent this state is not normalizable. This is
understandable since we have seen in eq (\ref{ancre}) that
$\bar{a}^-$ is in fact a creation operator, so the state
annihilated by it is in fact a maximal energy state instead of a
minimal energy state.

It is possible to represent the Fock vaccua in term of holomorphic
wave function if one allow the wave function to be  distributional in
that case the Fock
vaccua can be written as a product for each mode of
$  \delta(z^-)(k) \exp(z^+(k)z^+(k)/4|k|l_{P}^{2})$.

\section{linearized Kodama state}

The Kodama state $ \psi_{K}(A) = \exp(S_{CS}(A)
/\mathrm{H}^{2}l_{P}^{2})$, where \be S_{CS}(A) = \int \frac{1}{2}A \wedge
dA + \frac{1}{3}A\wedge A\wedge A \ee is the unique solution of
the quantum self-duality equation \be B^c_{k} \psi_{K}(A)  -
\mathrm{H}^{2}l_{P}^{2} \frac{\delta \psi_{K}}{\delta A_{c}^k}=0. \ee We
want to expand the Chern-Simons functional around de Sitter
background,  for symmetric transverse traceless perturbation one
obtains $S_{CS} = S_{0}+ S(a) + I(a)$  where \be S_0 =
-i\mathrm{H}^{3}\int f^{3}, \ee the quadratic fluctuation is given, up to
boundary terms, by \be S(a) =\frac{1}{2} \int
\epsilon^{abc}\partial_{a}a_{b}^ia_{c}^i -i \mathrm{H} f a_{b}^c a_{c}^b,
\ee and \be I(a) = \frac{1}{6}\int
d^{3}x\epsilon_{ijk}\epsilon^{abc} a_{a}^ia_{b}^ja_{c}^k. \ee It
is convenient to  express  the quadratic fluctuation in terms of
fourier modes $ S(a)= S^{+}(a^{+}) +S^-(a^-)$ with \be
S^{\pm}(a^{\pm}) = \frac{1}{2}  \int d^{3}k (\pm |k| -i\mathrm{H} f)
a^\pm(k)a^\pm(-k). \ee We are interested into small fluctuation
around the de Sitter background. The  linearized Kodama state
$\psi_{LK}$ is given by \be \psi_{LK}(f,z^{+},z^-) =
e^{S_{0}/\mathrm{H}^2l_{P}^{2}}e^{S^{+}(z^{+})/\mathrm{H}^{2}l_{P}^{2}
+S^-(z^-)/\mathrm{H}^{2}l_{P}^{2}}. \ee $S_{0}$ is infinite since it is
the integral of a constant $f$ which contains an infinite volume
factor. One way to deal with that is by cutting off the flat slice
at a fix volume $V$ and then let $V$ goes to infinity. The
linearized Kodama state is, as a function of $a$ and up to a
constant, the unique solution of the linearized self-duality
equation \be (\pm |k| -i\mathrm{H} f)a^\pm(k) \psi_{LK}(z^+,z^-)
-\mathrm{H}^2l_{P}^{2}  \frac{\partial \psi(z^+,z^-)}{\partial z^\pm(-k)}
=0, \ee and it is explicitly time dependent. This state is a good
approximation of the full Kodama state if we can neglect the cubic
term $I(a)$, this is the case  when the fluctuations satisfy $
|\int d^3q\, a^\pm(q-k)a^\pm(-q)|<<|ka^\pm(k)|$.

\subsection{scalar product}

The linearized Kodama state is given by  a product of an
holomorphic function for  positive helicity with an
holomorphic function for negative helicity. The measure of
integration of the scalar product has the same property, so the norm
of $\psi_{LK}$ factorizes
$||\psi_{LK}||^2= |\psi_{LK}|_{+}^2 |\psi_{LK}|_{-}^2$, with
\be
 |\psi_{LK}|_{\pm}^2 = \int {\cal{D}}^2z^\pm e^{ Q^\pm(z^\pm,\bar{z}^\pm)}.
\ee
$Q^\pm$ are quadratic forms given by
\be
\begin{array}{rcl}
Q^\pm(z^\pm,\bar{z}^\pm) &=&
\int d^{3}k   \left(
      \begin{array}{c}
             {z}^\pm(k)\\
             \bar{z}^\pm(-k)
      \end{array}
\right)^t
{\mathbf{Q}}^\pm(k)
\left(
      \begin{array}{c}
             \bar{z}^\pm(k)\\
             z^\pm(-k)
      \end{array}
\right), \\
 \\
{\mathbf{Q}}^\pm(k) &=& \frac{\mp1}{l_P^2} \left(
\begin{array}{cc}
    \frac{f^2}{4 |k|} &  \frac{f^2}{4 |k|} - \frac{|k|}{2\mathrm{H}^{2}}
    \pm i\frac{ f}{2\mathrm{H}}  \\
    \frac{f^2}{4 |k|} - \frac{|k|}{2\mathrm{H}^{2}} \mp i 
    \frac{ f}{2\mathrm{H}} &
    \frac{f^2}{4 |k|}
\end{array}\right)
\end{array}
\ee
In order to know the eigenvalues of $\mathbf{Q}$ lets compute
\beq
\mathrm{tr}{\mathbf{Q}}^\pm(k) &=&\mp \frac{f^2}{2 |k|l_{P}^{2}}, \\
\det{\mathbf{Q}}^\pm(k) & =& - \left(\frac{|k|}{2\mathrm{H}^2 
l_{P}^{2}}\right)^ 2.
\eeq
Since the determinant is always negative,
we see that for both helicities at least one of the eigenvalue is
positive so the corresponding mode
is not normalizable. This lead to the drastic conclusion that for both 
helicities there is always a non
normalizable mode.
In order to better understand the nature of this non normalisability of
the Kodama state at the quadratic level
we can decompose $a$ into real and imaginary part
$z^\pm(k) =x^\pm(k)+ i y^\pm(k)$, and $x,y$ are real in the sense that
$\bar{x}(k) =x(-k)$, $\bar{y}(k) =y(-k)$.
With this variables we can write the quadratic form as
\be
{\mathbf{Q}}^\pm(k)=\pm  \frac{|k|}{\mathrm{H}^2 l_P^2} \left\{
x^\pm(k)\bar{x}^\pm(k)
- [y^\pm(k)\mp \frac{\mathrm{H} f}{|k|} x^\pm(k)][\bar{y}^\pm(k)\mp 
\frac{\mathrm{H} f}{|k|} \bar{x}^\pm(k)]
\right\}
\ee
One clearly sees that the non normalisabilty to quadratic order of the
Kodama state is
due to non normalizable fluctuation for each mode $k$ and for each helicity.
This is
{ very different} in nature to the
non normalisability of the Chern-simons state in electromagnetism.
In the latter case we can show that the positive helicity modes are normalizable
whereas
the negative helicity modes are non normalizable. Also the positive helicity
modes have a positive energy while the negative
helicity modes have a negative energy.
The Kodama state do not show such drastic birefringence properties since
none of the helicity are
normalizable to quadratic order.

Finally, It is not clear if the instability (non-normalisability) of the
linearized Kodama state implies some instability
for the full Kodama state.
What we have proven so far is the non normalisability of the Kodama state
{\it to quadratic order}
we have neglected the influence of higher order correction.
The full Kodama state is cubic, so we do not expect the higher order
terms of the Chern-Simons term to improve the convergence
properties.
On the other hand the measure of integration implementing the reality conditions  will introduce contribution
{\it to all order}.
This is very different from the QCD case where there are non contribution
coming from the measure.
So there is still the logical possibility that the Kodama state while non
normalizable to quadratic order is
normalizable when we take into account the contribution from the measure
to all order.

\section{The linearized Euclidean theory}

It is easy to see that the linearized Kodama state is delta-functional
normalizable in the Euclidean theory.  The
reduction to the two helicity states goes the same way as the
Lorentzian theory. Hence the theory is again reduced to linearized
physical degrees of freedom, $(a^\pm (k), e^\pm (k))$. However these
are separately real, because the Euclideanized reality
conditions are simply that $E^{ai}$ and $A_{ai}$ are real.  There is
no $\imath$ in the classical Poisson brackets, so the canonical
commutation relations are now,
\be
[a^{\pm}(k),e^{\pm}(p)]=\imath l_P^2 \delta^3(k+p);\ \ \  [a^\pm(k),e^\mp(p)]=0.
\ee
The states in the linearized Hilbert space are again functionals
$\psi (z^+, z^-)$.
The representation is defined by
\be
  a^\pm(k) \psi(z^+,z^-)= z^\pm(k) \psi(z^+,z^-).
  \ee
The representation we work with is the one for which
the frame fields acts as a derivative operator
  \be
  e^\pm(-k) \psi(z^+,z^-) =
  -\imath l_{P}^{2} \frac{\partial \psi(z^+,z^-)}{\partial z^\pm(k)}.
  \ee
The inner product that realizes the reality conditions is now simply
\be
<\phi|\psi> = \int {\cal{D}}^2z^+ {\cal{D}}^2z^-
\overline{\phi}(\bar{z}^+,\bar{z}^-)  \psi(z^+,z^-),
\ee
where the integration is over a real section $ z(k)=\bar{z}(-k)$.
The linearized self-duality condition on states is now
\be
(\pm |k| -\mathrm{H} f)a^\pm(k) \psi_{LK}(z^+,z^-) -\imath \mathrm{H}^2l_P^2  \frac{\partial
\psi(z^+,z^-)}{\partial z^\pm(-k)} =0,
\ee
The $\imath$ in the last term is now there because there is an
$\imath$ in the canonical commutation relation. An $\imath$ that is in
the first term in the Lorentzian theory is absent because the
connection $A_a^i$ corresponding to Euclidean deSitter is purely real
rather than purely imaginary.

The unique solution to the linearized self-dual equation is again
a linearized Kodama state. It is now
\be
\psi_{ELK}(f,z^{+},z^-) = e^{\imath S_{0}/\mathrm{H}^2l_P^2}e^{\imath
S^{+}(z^{+})/\mathrm{H}^{2}l_P^2 +\imath S^-(z^-)/\mathrm{H}^{2}l_P^2}.
\ee
where $S_0$ and the $S^\pm$ are now real functionals of $z^\pm$.

The result is that the norm of the linearized Euclidean Kodama state is
\be
<ELK|ELK> = \int {\cal{D}}^2z^+ {\cal{D}}^2z^-  1
\ee
This is delta functional normalizable.

\section{Alternatives}

Before closing we want to make some comment on the implications of
the results we described here.

We first may note that the issue or normalizability would not
generally come up for a semiclassical state of wkb form for some
system
\f
\Psi (x) \approx e^{\imath S(x)^{HJ}}
\ff
where $S(x)^{HJ}$ is a solution to the Hamiltonian-Jacobi functions.
Such states are only delta-function normalizable. As energy is one of
the parameters of the Hamilton-Jacobi equation, such solutions correspond
to a definite choice of energy. Normalizable states
are wavepackets constructed from superposition of
energy eigenstates.
This is not generally a problem for
normal systems. The problem for us is that on
compact regions, and in the absence of matter, the Kodama state is unique.
There is no parameter to vary, it depends only on
$\lambda=G\hbar\Lambda$ which is a parameter of the theory

One possibility is to couple gravity to matter with a potential,
such that the value of the cosmological constant becomes a
parameter of a solution. This case one can consider wavepackets
constructed by superposing different values of $\lambda$. Such a
procedure has been recently proven successful
\cite{Alexander:2003wb} in the case of mini-superspace
quantization of gravity couple to a scalar field which involves
only the zero modes of the Kodama state. Since the Euclidean
linearized Kodama state is only delta functional normalisable this
procedure is expected to be successful, but only in the Euclidean
case. It is not known whether or not it works in the full theory.

There is another simple consideration  that shed some light on the
meaning of the truncation of the Kodama state. We earlier wrote
the Kodama state in the form \f \Psi_K (A) = e^{S^0 + S^2 + S^3}
\ff Truncation corresponded to dropping the $S^3$ term. In the
ordering in which the full Kodama state is a solution to the
constraints we may write the Hamiltonian constraint schematically
as \f H= EEJ \ff where $J= F+ H^2 E$ is the self-duality operator.
Each has an expansion around deSitter spacetime. \f J=J^1 + J^2 ;
\ \ \ \ E= E^0 +e \ff We can then write \f H= H^1 + H^2 +H^3 + H^4
\ff As we indicated the truncated state $\Psi_{LK}=e^{S^0+S^2}$ is
a physical state of the linearized theory \f H^1 \Psi_{LK} =0 \ff
Moreover, the linearized Kodama  is explicitly time dependent, The
quadratic hamiltonian is also time dependent its dynamics is
govern by the quadratic Hamiltonian. One can check that \f H^2
\Psi_{LK} = \frac{\partial}{\partial \eta} \Psi_{LK} \neq 0. \ff
The linear Kodama state thus possess all the properties we would
expect for a vaccua except that this is not a Fock vaccua.

The main question is then wether this means  that the full
Kodama state cannot be physical if its linearization is not a
normalizable state in the inner product of the linearized theory.
Since one should  expect that if one has a state which
is proportional to be the ground state of the full theory, its truncation
to a linearized state should be the ground state of the linearized
theory.  And, certainly, as we have seen the linearized Kodama state is not
the ground state of the linearized theory.

Let us try to flesh out this argument to see how definitive it is.
To state it more precisely requires that for at least some classical
solutions, $A^0,E_{0}$ there exists a map $\cal M$
from the full to linearized physical Hilbert spaces,
\f
{\cal M}: {\cal H}^{physical} \rightarrow {\cal H}^{linearized}
\ff
satisfying some natural list of properties.  What should these be,

\begin{enumerate}

    \item{}{\cal M} is defined for solutions $A^0,E_{0}$ of maximum symmetry

    \item{}The image of the ground state is the ground state

    \item{}There is a subspace ${\cal H}^{full gravitons} \subset
    {\cal H}^{physical}$ which is mapped into
    ${\cal H}^{linearized}$.  This corresponds to gravitons propagating on
    the  background, fully dressed in the fully interacting theory.

    \item{}$\cal M$ takes states that are fully diffeomorphism
    invariant to states that are invariant under linearized
    diffeomorphisms around $A^0$.

    \item{}The orthogonal subspace to ${\cal H}^{full gravitons}$ is
    mapped to the null vector in ${\cal H}^{linearized}$. Hence there
    is a large kernel.  These correspond to states in ${\cal
    H}^{physical}$ that cannot be decomposed in the basis of graviton
    states on the given background.

    \item{}If we accept the results in loop quantum gravity indicating
    there is an ultraviolet cutoff, such as the discreteness of area
    and volume, then the map $\cal M$ cannot be onto, because there
    will be no states in its image with wavelength or frequency
    shorter than $l_{Planck}.$
    \footnote{We note that this implies that the symmetry group of $A^0$ is
    either broken or deformed as in doubly special 
    relativity \cite{Amelino-Camelia:2000mn}.
    We note that it has recently been established that the latter is
    the case in $2+1$ gravity \cite{Freidel:2003sp}.}

    \item{}We note that the choice of a maximally symmetric Lorentzian
    spacetime $A^0$ does not determine a unique Hilbert space of
    linearized fields. Additional information is required
    corresponding to the choice of a timelike killing field on all or
    part of $A^0$.  The Hilbert spaces corresponding to different
    choices  are generally unitarily inequivalent. Examples are the
    Minkowski vrs the Rindler states in Minkowski spacetime or the
    Hilbert spaces corresponding to different observers in deSitter
    spacetime. Thus,  $\cal M$ must depend on additional information
    beyond the specification of $A^0$.

\end{enumerate}

We may note that if we insist on properties $4$ to $7$ the map may not be
just a simple
truncation of the functional form. Hence, it may be not necessary that
${\cal M} \cdot \Psi_K \rightarrow \Psi_{LK}$.

Another problem with such a map $\cal M$  is contained in
property $2$  The problem is that there is no definition
of the Hamiltonian for the full theory, in the absence of a boundary.
In classical or quantum gravity, it is a simple and direct consequence of the
equivalence principle that energy is only defined quasi-locally, on a
timelike or null $3$-surface
which may be taken to be the boundary of a region of spacetime. The
boundary may be at infinity, as in the $ADM$ energy or it may be of
finite area. But without a
boundary there can be no definition of energy.

Finally, It is well known that the usual choice of Fock vaccua is
well defined only after one has given a choice of synchronized
observers, for instance by specifying a timelike Killing vector field on
the background.
The Kodama state is a full covariant state, in that it does not
depend on the choice of a timelike killing field on a background. It
cannot, for it is defined on {\it any} point in the configuration space.
A point in the configuration space corresponds to a connection on a
three slice, hence it is dual to a $3$-geometry, not a solution.
Further, only a set of measure zero correspond to spatial slices of metrics that
have killing fields.

A choice of a time like killing field corresponds in some sense to
a choice of an observer in spacetime.  We may then try to
interpret the fact that the Kodama state does not map to a
linearized vacua as saying that the linearized vacua depends on a
choice of an observer which is not made in the specification of
the Kodama state. It either means that the Kodama state is not
physical because its covariance prevent its linearization from
being a Fock vaccua or that the map $\cal M$ should also contain
in some way a choice of synchronized  observer.

Thus, the conclusion is that while the lack of
normalizability of the linearization of the Lorentzian Kodama
state is worrying, there is not yet a definitive argument that
the full state is unphysical in the Lorentzian case. We also do not
yet understand the significance of the fact that the Euclidean version
of the Kodama state appears to be better off, in this respect.
More work is clearly needed.   Among the issues left open are
the question of how to evaluate the action of the
spin foam projection operator onto
physical states on the Kodama state. Also,
to be studied in future work, is a further analysis of  the
hypothesis that the physics in the presence of the Kodama state
reproduces in an appropriate limit the quantization of field in a de Sitter
background.

\section{Acknowledgments}\nonumber

We would like to thank E. Witten for correspondance, C. Rovelli
for being a collaborator in the beginning of the project,
 S. Alexander, M. Bojowald, D. Minic,  A. Starodubtsev, and especially
R. Myers for discussions.  We also thank Roman Jackiw for pointing out
a silly error in the discussion of Yang-Mills theory in 
an earlier version of this paper.

\end{document}